\documentclass[prb,preprint]{revtex4-1} 

\usepackage{amsmath}  
\usepackage{amsfonts} 
\usepackage{graphicx} 

\DeclareMathOperator{\dd}{d\!}

\begin{document}

\title{Demystifying the Lagrangian formalism for field theories}

\author{Gerd Wagner}
\email{gerdhwagner@t-online.de} 
\affiliation{Mayener Str. 131, 56070 Koblenz, Germany} 

\author{Matthew W. Guthrie}
\email{matthew.guthrie@ucf.edu}
\affiliation{Department of Physics, University of Central Florida, Orlando, FL 32816}


\date{\today}

\begin{abstract} This paper expands on previous work\cite{guthrie2019demystifying} to derive and motivate the Lagrangian formulation of field theories.
In the process, we take three deliberate steps.
First, we give the definition of the action and derive Euler-Lagrange equations for field theories.
Second, we prove the Euler-Lagrange equations are independent under arbitrary coordinate transformations and motivate that this independence is desirable for field theories in physics.
We then use the Lagrangian for Electrodynamics as an example field Lagrangian and prove that the related Euler-Lagrange equations lead to Maxwell's equations.
\end{abstract}

\maketitle

\section{Introduction}

When Lagrangian field theory is introduced it is often presented as a generalization of the Lagrangian formalism of classical mechanics, and this profoundly shapes physicists' understanding of the subject.
For the most common modern example, see Goldstein's treatment of classical field theories \cite{goldstein2002classical}, which appears in one of the common graduate texts in classical mechanics.
The approach of this paper differs from traditional approaches in that we start by presenting the Lagrangian formulation of field theories as a purely mathematical formalism.
We find that the Lagrangian formulation has very well defined coordinate and field transformation properties.
Because we consider these properties extremely useful for physical field theories, the desire to find Lagrangians for these theories in order to turn their field equations into Euler-Lagrange equations is well motivated.
As a proof of concept, we provide the Lagrangian for Electrodynamics as a definition and show that it leads to Maxwell's equations.
The domain of this work is purely non-relativistic which means neither the Lagrangian formulation of the field theories nor the treatment of Electrodynamics requires invoking concepts from special relativity. 

\section{Definition of the Lagrangian formalism for fields}\label{definition}

The experienced reader may recognize the symbols and names used in this section.
Nonetheless, this section should be thought of as containing purely mathematical definitions and conclusions.
We define a field as a function $\psi(t,x)$ of time $t$ and of three spatial coordinates denoted by $x$. 
The field's values may be multidimensional.
A well-known example is the electric field, which has a direction in space and whose values are thus three-dimensional.

A Lagrangian $\mathcal{L}$ of $\psi$ is defined as a function that may depend on $\psi$ itself as well as on its time and spatial derivatives: 
\begin{equation}
\mathcal{L} = \mathcal{L}\left(\psi, \frac{\partial \psi}{\partial t}, \frac{\partial \psi}{\partial x}\right).
\end{equation}
The action $S$ for two points $t_1$ and $t_2$ in time and a three dimensional area of space $A$ is defined as the following integral of the Lagrangian:
\begin{equation}
S := \int\limits_{t_1}^{t_2} \int_{A} \mathcal{L}\left(\psi, \frac{\partial \psi}{\partial t}, \frac{\partial \psi}{\partial x}\right) \dd x^3 \dd t.
\end{equation}

Next, we are interested in the conditions that $\mathcal{L}$ must fulfill to make $S$ stationary.
To do so we consider arbitrary but small variations $\delta\psi$ of the field and calculate the resulting variation $\delta S$ of $S$. We require the variations $\delta\psi$ to vanish at $t_1$ and $t_2$ as well as on the surface of $A$,
\begin{equation}\label{integrands}
\delta S = \int\limits_{t_1}^{t_2} \int_{A} 
\frac{\partial \mathcal{L}}{\partial \psi} \cdot \delta \psi
+ \frac{\partial \mathcal{L}}{\partial \frac{\partial \psi}{\partial t}} \; \delta \left(\frac{\partial \psi} {\partial t}\right)
+ \frac{\partial \mathcal{L}}{\partial \frac{\partial \psi}{\partial x}}  \; \delta \left(\frac{\partial \psi} {\partial x}\right)
\dd x^3 \dd t.
\end{equation}
If we consider the possibly multidimensional values of $\psi$ indexed by $j$ and the three spacial dimensions indexed by $i$, the integrands in equation \eqref{integrands} can be rewritten as:
\begin{align}
\frac{\partial \mathcal{L}}{\partial \psi} \; \delta \psi
&= \sum_{j} \frac{\partial \mathcal{L}}{\partial \psi_{j}} \; \delta \psi_{j}, \\
\frac{\partial \mathcal{L}}{\partial \frac{\partial \psi}{\partial t}} \; \delta \left(\frac{\partial \psi} {\partial t}\right)
&= \sum_{j} \frac{\partial \mathcal{L}}{\partial \frac{\partial \psi_{j}}{\partial t}} \; \delta \left(\frac{\partial \psi_{j}} {\partial t}\right), \\
\frac{\partial \mathcal{L}}{\partial \frac{\partial \psi}{\partial x}} \; \delta \left(\frac{\partial \psi} {\partial x}\right)
&= \sum_{i,j} \frac{\partial \mathcal{L}}{\partial \frac{\partial \psi_{j}}{\partial x_{i}}} \; \delta \left(\frac{\partial \psi_{j}} {\partial x_{i}}\right).
\end{align}

We now integrate the second summand by parts over time and the third summand by parts over space.
To do so, we use the identities
$\delta \left(\frac{\partial \psi} {\partial t}\right)
= \frac{\partial \psi_2} {\partial t} - \frac{\partial \psi_1} {\partial t}
= \frac{\partial (\psi_2 - \psi_1)} {\partial t}
= \frac{\partial \delta \psi} {\partial t}$ 
and
$\delta \left(\frac{\partial \psi} {\partial x}\right)
= \frac{\partial \psi_2} {\partial x} - \frac{\partial \psi_1} {\partial x}
= \frac{\partial (\psi_2 - \psi_1)} {\partial x}
= \frac{\partial \delta \psi} {\partial x}$.
The integral is performed as follows:
\begin{equation} \label{calcDeltaSSection2}
\begin{split}
\delta S = \int\limits_{t_1}^{t_2} \int_{A} 
\frac{\partial \mathcal{L}}{\partial \psi} \delta \psi
-\left(\frac{\partial}{\partial t} \left( \frac{\partial \mathcal{L}}{\partial \frac{\partial \psi}{\partial t}} \right)\right) \delta \psi
-\left(\frac{\partial}{\partial x} \cdot \left( \frac{\partial \mathcal{L}}{\partial \frac{\partial \psi}{\partial x}} \right)\right) \delta \psi \;
\dd x^3 \dd t \\
+ \int_{A} \int\limits_{t_1}^{t_2} \frac{\partial}{\partial t} \left(\frac{\partial \mathcal{L}}{\partial \frac{\partial \psi}{\partial t}} \; \delta \psi \right) \dd t \dd x^3
+ \int\limits_{t_1}^{t_2} 
\int_{A} \frac{\partial}{\partial x} \cdot \left( \frac{\partial \mathcal{L}}{\partial \frac{\partial \psi}{\partial x}} \; \delta \psi \right) \dd x^3  \dd t.
\end{split}
\end{equation}
with ``$\frac{\partial}{\partial x} \cdot$'' denoting the divergence with respect to the coordinates $x$ (we refrain from using the usual ``$\nabla \cdot$'' because later the divergence with respect to variables other than $x$ will occur).

The second integral vanishes as a result of the fundamental theorem of calculus and because $\delta \psi(t_1) = \delta \psi(t_2) = 0$.
The third integral vanishes from use of Gauss's theorem and because $\delta \psi(x) = 0$ for any $x$ on the surface of $A$. Thus,
\begin{equation}
\delta S = \int\limits_{t_1}^{t_2} \int_{A} 
\frac{\partial \mathcal{L}}{\partial \psi} \delta \psi
-\left(\frac{\partial}{\partial t} \left( \frac{\partial \mathcal{L}}{\partial \frac{\partial \psi}{\partial t}} \right)\right) \delta \psi
-\left(\frac{\partial}{\partial x} \cdot \left( \frac{\partial \mathcal{L}}{\partial \frac{\partial \psi}{\partial x}} \right)\right) \delta \psi \;\;
\dd x^3 \dd t.
\end{equation}
If we use the same index conventions for the possibly multidimensional values of $\psi$ and the three spacial dimensions as we did above, the last two summands of the integrand become:
\begin{align}
\frac{\partial}{\partial t} \left( \frac{\partial \mathcal{L}}{\partial \frac{\partial \psi}{\partial t}} \right) \; \delta \psi
&= \sum_j \frac{\partial}{\partial t} \left( \frac{\partial \mathcal{L}}{\partial \frac{\partial \psi_j}{\partial t}} \right) \; \delta \psi_j \\
\left(\frac{\partial}{\partial x} \cdot \left( \frac{\partial \mathcal{L}}{\partial \frac{\partial \psi}{\partial x}} \right)\right) \; \delta \psi
&= \sum_j \left(\sum_i \frac{\partial}{\partial x_i} \; \left( \frac{\partial \mathcal{L}}{\partial \frac{\partial \psi_j}{\partial x_i}} \right)\right) \; \delta \psi_j.
\end{align}

The last rewrite of $\delta S$ we wish to do is
\begin{equation}
\delta S = \int\limits_{t_1}^{t_2} \int_{A} 
\left(
\frac{\partial \mathcal{L}}{\partial \psi}
-\frac{\partial}{\partial t} \left( \frac{\partial \mathcal{L}}{\partial \frac{\partial \psi}{\partial t}} \right) 
-\frac{\partial}{\partial x} \cdot \left( \frac{\partial \mathcal{L}}{\partial \frac{\partial \psi}{\partial x}} \right)\right) \delta \psi \;
\dd x^3 \dd t.
\end{equation}
Because $\delta \psi$ is arbitrary except for its boundary conditions, the only way to make $S$ stationary (which is equivalent to requiring that $\delta S = 0$) is that $\mathcal{L}$ fulfills the condition

\begin{equation}
0 = \frac{\partial \mathcal{L}}{\partial \psi}
-\frac{\partial}{\partial t} \left( \frac{\partial \mathcal{L}}{\partial \frac{\partial \psi}{\partial t}} \right) 
-\frac{\partial}{\partial x} \cdot \left( \frac{\partial \mathcal{L}}{\partial \frac{\partial \psi}{\partial x}} \right).
\end{equation}

This is the Euler-Lagrange equation for field theory.
It can be seen as a counterpart of the Euler-Lagrange equation for classical particles discussed in previous work \cite{guthrie2019demystifying}.
The procedure of looking for a condition to make $S$ stationary under a Lagrangian $\mathcal{L}\left(\psi, \frac{\partial \psi}{\partial t}, \frac{\partial \psi}{\partial x}\right)$ is called the Lagrangian formalism for field theory.

\section{Invariance of the Euler-Lagrange equation for field theory under transformations} \label{sectionInvariance}

Let $x=f(\bar{x})$ be an invertible and differentiable transformation of the spatial coordinates and $\psi=F(\bar{\psi})$ be an invertible and differentiable transformation of the field (for a discussion of this transformation see appendix \ref{appendixDiscussionTransform}).
We define the transformed Lagrangian  $\bar{\mathcal{L}}$ as

\begin{equation} \label{LagrTransform}
\bar{\mathcal{L}}\left(\bar{\psi}, \frac{\partial \bar{\psi}}{\partial t}, \frac{\partial \bar{\psi}}{\partial \bar{x}}\right) 
:= \mathcal{L}\left(F(\bar{\psi}), \frac{\partial F(\bar{\psi})}{\partial t}, \frac{\partial F(\bar{\psi})}{\partial f}\right) 
\left| \det \frac{\partial f}{\partial \bar{x}} \right|,
\end{equation}
where $\left| \det \frac{\partial f}{\partial \bar{x}} \right|$ is the absolute value of the determinant of the Jacobian matrix of $f$ with respect to the spatial coordinates $\bar{x}$. \\

We will prove that, by requiring $S$ to be stationary, the equations

\begin{equation} \label{ELGTransformed}
0 = \frac{\partial \bar{\mathcal{L}}}{\partial \bar{\psi}}
-\frac{\partial}{\partial t} \left( \frac{\partial \mathcal{\bar{L}}}{\partial \frac{\partial \bar{\psi}}{\partial t}} \right) 
-\frac{\partial}{\partial \bar{x}} \cdot \left( \frac{\partial \mathcal{\bar{L}}}{\partial \frac{\partial \bar{\psi}}{\partial \bar{x}}} \right) 
\end{equation}
and
\begin{equation} \label{ELGUntransformed}
0 = \frac{\partial \mathcal{L}}{\partial \psi}
-\frac{\partial}{\partial t} \left( \frac{\partial \mathcal{L}}{\partial \frac{\partial \psi}{\partial t}} \right) 
-\frac{\partial}{\partial x} \cdot \left( \frac{\partial \mathcal{L}}{\partial \frac{\partial \psi}{\partial x}} \right) 
\end{equation}
follow, and thus that the Euler-Lagrange equation for field theory is independent of arbitrary coordinate and field transformations as long as the transformation of the Lagrangian is given by equation \eqref{LagrTransform}. \\

To do so, we consider arbitrary but small variations $\delta \bar{\psi}$ of the field $\bar{\psi}$ that vanish at $t_1$, $t_2$, and on the surface of an area of space $\bar{A}$. We use these to find the condition for  
\begin{equation}
S = \int\limits_{t_1}^{t_2} \int_{\bar{A}} \bar{\mathcal{L}}\left(\bar{\psi}, \frac{\partial \bar{\psi}}{\partial t}, \frac{\partial \bar{\psi}}{\partial \bar{x}}\right) \dd \bar{x}^3 \dd t 
= \int\limits_{t_1}^{t_2} \int_{\bar{A}} \mathcal{L}\left(F(\bar{\psi}), \frac{\partial F(\bar{\psi})}{\partial t}, \frac{\partial F(\bar{\psi})}{\partial f}\right) 
\left| \det \frac{\partial f}{\partial \bar{x}} \right| \dd \bar{x}^3 \dd t 
\end{equation}
to become stationary. 
Equation \eqref{ELGTransformed} follows from repeating the considerations of section \ref{definition}.
To prove equation \eqref{ELGUntransformed}, we look at
\begin{equation}
S = \int\limits_{t_1}^{t_2} \int_{\bar{A}} \mathcal{L}\left(F(\bar{\psi}), \frac{\partial F(\bar{\psi})}{\partial t}, \frac{\partial F(\bar{\psi})}{\partial f}\right) 
\left| \det \frac{\partial f}{\partial \bar{x}} \right| \dd \bar{x}^3 \dd t 
\end{equation}
which, using the transformation formula of multidimensional integrals (see Appendix \ref{TranformationFormula}), can be transformed into
\begin{equation}
S = \int\limits_{t_1}^{t_2} \int_{f(\bar{A})} \mathcal{L}\left(F(\bar{\psi}), \frac{\partial F(\bar{\psi})}{\partial t}, \frac{\partial F(\bar{\psi})}{\partial f}\right) 
\dd f^3 \dd t,
\end{equation}
where $f(\bar{A})$ is the representation of $\bar{A}$ under the coordinate transformation $f$.
Based on this formula, the variation $\delta S$ of $S$ is given by
\begin{equation}
\delta S = \int\limits_{t_1}^{t_2} \int_{f(\bar{A})} 
\frac{\partial \mathcal{L}}{\partial F} \; \delta F
+ \frac{\partial \mathcal{L}}{\partial \frac{\partial F}{\partial t}} \; \delta \left(\frac{\partial F}{\partial t}\right)
+ \frac{\partial \mathcal{L}}{\partial \frac{\partial F}{\partial f}} \; \delta \left(\frac{\partial F} {\partial f}\right)
\dd f^3 \dd t,
\end{equation}
where 
\begin{equation} \label{deltaFDefinition}
\delta F = \frac{\partial F}{\partial \bar{\psi}} \delta \bar{\psi}.
\end{equation}
Integration by parts of the second and the third term leads to
\begin{equation} \label{calcDeltaSSection3}
\begin{split}
\delta S = \int\limits_{t_1}^{t_2} \int_{f(\bar{A})} 
\frac{\partial \mathcal{L}}{\partial F} \delta F
-\left(\frac{\partial}{\partial t} \left( \frac{\partial \mathcal{L}}{\partial \frac{\partial F}{\partial t}} \right)\right) \delta F
-\left(\frac{\partial}{\partial f} \cdot \left( \frac{\partial \mathcal{L}}{\partial \frac{\partial F}{\partial f}} \right)\right) \delta F \;
\dd f^3 \dd t \\
+ \int_{f(\bar{A})} \int\limits_{t_1}^{t_2} \frac{\partial}{\partial t} \left(\frac{\partial \mathcal{L}}{\partial \frac{\partial F}{\partial t}} \; \delta F \right) \dd t \dd f^3
+ \int\limits_{t_1}^{t_2} 
\int_{f(\bar{A})} \frac{\partial}{\partial f} \cdot \left( \frac{\partial \mathcal{L}}{\partial \frac{\partial F}{\partial f}} \; \delta F \right) \dd f^3 \dd t,
\end{split}
\end{equation}
where the identities $\delta \left(\frac{\partial F} {\partial t}\right)
= \frac{\partial F_2} {\partial t} - \frac{\partial F_1} {\partial t}
= \frac{\partial (F_2 - F_1)} {\partial t}
= \frac{\partial \delta F} {\partial t}$ 
and
$\delta \left(\frac{\partial F} {\partial f}\right)
= \frac{\partial F_2} {\partial f} - \frac{\partial F_1} {\partial f}
= \frac{\partial (F_2 - F_1)} {\partial f}
= \frac{\partial \delta F} {\partial f}$ 
were used.
Following from equation \eqref{deltaFDefinition}, $\delta F$ vanishes at $t_1$ and $t_2$ as $\delta \bar{\psi}$ does, and the second term in \eqref{calcDeltaSSection3} is zero because of the fundamental theorem of calculus.
Using Gauss's theorem, the third term can be transformed into an integral over the surface of $f(\bar{A})$ which we denote by $\partial (f(\bar{A}))$.
This surface is the same as the representation of the surface of $\bar{A}$ under $f$:
\begin{equation} \label{surfaceEquality}
\partial (f(\bar{A})) = f(\partial \bar{A}).
\footnote{The simplest way to picture this equation is to imagine a real area in space, which is described from within two systems of coordinates.}
\end{equation}
To show that the third term vanishes, we will prove that $\delta F$ is zero for any $x \in \partial (f(\bar{A}))$.

First, let $x$ be an element of $\partial (f(\bar{A}))$.
\footnote{The simplest way to picture this element is to imagine a real point on the surface of the area in space, which is described from within two systems of coordinates.}
Then for $x$ there exists a unique $\bar{x} \in \partial \bar{A}$ which is defined by $x=f(\bar{x})$. We use the fact from above that $\delta \bar{\psi}(\bar{x}) = 0$ and recall that the variation $\delta \bar{\psi}$ is a difference between two fields (which we denote $\bar{\psi}_1$ and $\bar{\psi}_2$) such that
\begin{equation}
\delta \bar{\psi} = \bar{\psi}_2 - \bar{\psi}_1.
\end{equation}
The value of $F$ considered as a function of $x$ is given by 
\begin{equation} \label{defineFOfx}
F(x) = F(\bar{\psi}(\bar{x})) \;\; \text{with} \;\; \bar{x} \;\; \text{defined through} \;\; x=f(\bar{x}) 
\iff \bar{x} = f^{-1}(x).
\end{equation}
For a discussion of formula \eqref{defineFOfx}, see appendix \ref{appendixDiscussionTransform}.
The variation $\delta F$ that results from the difference $\delta \bar{\psi}$ between $\bar{\psi}_1$ and $\bar{\psi}_2$ is given by
\begin{equation}
\delta F(x) = F(\bar{\psi}_2(\bar{x})) - F(\bar{\psi}_1(\bar{x})) 
= F(\bar{\psi}_1(\bar{x}) + \delta \bar{\psi} (\bar{x})) - F(\bar{\psi}_1(\bar{x})) 
= \frac{\partial F}{\partial \bar{\psi}} \delta \bar{\psi} (\bar{x}).
\end{equation}
Because $\delta \bar{\psi}(\bar{x})$ is zero by assumption, $\delta F(x)$ is zero, too, which finishes the proof.

As for $\delta S$, we are now left with
\begin{equation}
\delta S = \int\limits_{t_1}^{t_2} \int_{f(\bar{A})} 
\left(
\frac{\partial \mathcal{L}}{\partial F}
-\frac{\partial}{\partial t} \left( \frac{\partial \mathcal{L}}{\partial \frac{\partial F}{\partial t}} \right) 
-\frac{\partial}{\partial f} \cdot \left( \frac{\partial \mathcal{L}}{\partial \frac{\partial F}{\partial f}} \right)\right) \delta F \;\;
\dd f^3 \dd t.
\end{equation}
Because of equation \eqref{deltaFDefinition}, $\delta F$ is as arbitrary as $\delta \bar{\psi}$. Thus, the only way for $\delta S$ to vanish is
\begin{equation} \label{lastEqToProveInvariance}
0 = 
\frac{\partial \mathcal{L}}{\partial F}
-\frac{\partial}{\partial t} \left( \frac{\partial \mathcal{L}}{\partial \frac{\partial F}{\partial t}} \right) 
-\frac{\partial}{\partial f} \cdot \left( \frac{\partial \mathcal{L}}{\partial \frac{\partial F}{\partial f}} \right).
\end{equation}
If we now replace $F$ and $f$ according to their definitions by $\psi$ and $x$ this equation transforms into equation \eqref{ELGUntransformed} and thus finishes the proof.

\section{Application to physics: The Lagrangian for Electrodynamics}
The transformation properties for the Euler-Lagrange equations for field theories that we found on a purely mathematical basis makes the Euler-Lagrange formalism for field theories desirable for physics.
To make the formalism useful for physics we must turn the physical field equations into Euler-Lagrange equations.
This first requires finding a Lagrangian for the physical field theory in question.
This has been done for many existing physical field theories, such as Electrodynamics, General relativity, Schr\"odinger's equation, Dirac's equation, the Klein-Gordon equation, and the Standard model of particle physics.

As an example, we define the Lagrangian of Electrodynamics and show that Maxwell's equations can be derived from its Euler-Lagrange equations.
We start with some remarks on Maxwell's equations that can be found in greater detail in several textbooks, Jackson\cite{jackson1999classical} or Griffiths\cite{griffiths2017introduction} being two common examples.

With
\begin{itemize}
  \item $E$ denoting the three components of the electric field,
  \item $B$ denoting the three components of the magnetic field,
  \item $j$ denoting the three components of the electric current density,
  \item $\rho$ denoting the electric charge density,
  \item $\mu_0$ denoting permeability constant of empty space,
  \item $\epsilon_0$ denoting dielectric constant of empty space,
  \item $c$ denoting the speed of light,
\end{itemize}
Maxwell's equations in empty space are given by
\begin{align}
\nabla \times E &= - \frac{\partial B}{\partial t}, \label{Faraday} \\
\nabla \times B &= \mu_0 j +  \frac{1}{c^2} \frac{\partial E}{\partial t}, \\
\nabla \cdot E &= \frac{\rho}{\epsilon_0}, \\
\nabla \cdot B &= 0, \label{noMonopole}
\end{align}
where
\begin{equation}
\mu_0 = \frac{1}{\epsilon_0 c^2}.
\end{equation}
Because $\nabla \cdot B = 0$, there exists some vector potential $A$ such that
\begin{equation} \label{BbyA}
B = \nabla \times A.
\end{equation}
With that, equation \eqref{Faraday} can be rewritten as
\begin{equation}
\nabla \times \left( E + \frac{\partial A}{\partial t} \right) = 0.
\end{equation}
When the curl of a field is zero, the field can be expressed by the gradient of a scalar potential $\phi$. Thus, we can write
\begin{equation} \label{EbyAPhi}
E + \frac{\partial A}{\partial t} = - \nabla \phi
\iff
E = - \nabla \phi - \frac{\partial A}{\partial t}.
\end{equation}
To find equations \eqref{BbyA} and \eqref{EbyAPhi}, we refer to equations \eqref{noMonopole} and \eqref{Faraday}.
We now use equations \eqref{BbyA} and \eqref{EbyAPhi} to express the remaining two Maxwell equations using only $A$ and $\phi$:
\begin{align}
\nabla \times ( \nabla \times A) &= \mu_0 j + \frac{1}{c^2} \frac{\partial}{\partial t} \left(- \nabla \phi - \frac{\partial A}{\partial t} \right) \label{MaxwellAPhi1} \\
\nabla \cdot \left( -\nabla \phi - \frac{\partial A}{\partial t} \right) &= \frac{\rho}{\epsilon_0} . \label{MaxwellAPhi2}
\end{align}
We assert that these equations are the Euler-Lagrange equations of the Lagrangian
\begin{equation}
\mathcal{L} = \epsilon_0 \frac{E^2 - c^2 B^2}{2} - \rho\phi + j \cdot A
= \epsilon_0 \frac{(-\nabla\phi - \frac{\partial A}{\partial t})^2 - c^2 (\nabla \times A)^2}{2} - \rho\phi + j \cdot A.
\end{equation}
To prove this assertion, we first calculate the Euler-Lagrange equation for $\phi$
\begin{equation} \label{ELPhi}
0 = \frac{\partial \mathcal{L}}{\partial \phi}
-\frac{\partial}{\partial t} \left( \frac{\partial \mathcal{L}}{\partial \frac{\partial \phi}{\partial t}} \right) 
-\frac{\partial}{\partial x} \cdot \left( \frac{\partial \mathcal{L}}{\partial \frac{\partial \phi}{\partial x}} \right).
\end{equation}
We calculate the terms separately,
\begin{align}
\frac{\partial \mathcal{L}}{\partial \phi} &= -\rho \\
\frac{\partial \mathcal{L}}{\partial \frac{\partial \phi}{\partial t}} &= 0
\implies \frac{\partial}{\partial t} \left( \frac{\partial \mathcal{L}}{\partial \frac{\partial \phi}{\partial t}}\right) = 0 \\
\frac{\partial \mathcal{L}}{\partial \frac{\partial \phi}{\partial x}}
&= \frac{\epsilon_0}{2} \cdot 2 \left(-\nabla \phi - \frac{\partial A}{\partial t} \right) (-1)
= \epsilon_0 \left(\nabla \phi + \frac{\partial A}{\partial t} \right) \\
\frac{\partial}{\partial x} \cdot \left(\frac{\partial \mathcal{L}}{\partial \frac{\partial \phi}{\partial x}} \right)
&= \nabla \cdot \left[\epsilon_0 \left(\nabla \phi + \frac{\partial A}{\partial t} \right)\right].
\end{align}
Substituting these results into equation \eqref{ELPhi} results in
\begin{equation}
0 = -\rho - 0 -\nabla \cdot \left[\epsilon_0 \left(\nabla \phi + \frac{\partial A}{\partial t} \right)\right],
\end{equation}
which is equivalent to equation \eqref{MaxwellAPhi2}.
With that, the first part of the proof is done.

For the second part we must prove that
\begin{equation} \label{ELA}
0 = \frac{\partial \mathcal{L}}{\partial A}
-\frac{\partial}{\partial t} \left( \frac{\partial \mathcal{L}}{\partial \frac{\partial A}{\partial t}} \right) 
-\frac{\partial}{\partial x} \cdot \left( \frac{\partial \mathcal{L}}{\partial \frac{\partial A}{\partial x}} \right)
\end{equation}
is equivalent to equation \eqref{MaxwellAPhi1}.
We do this for the first component $A_1$ of $A$ only (note that equation \eqref{ELA} actually represents one equation for each component of $A$).
Again, we calculate the terms separately:
\begin{align}
\frac{\partial \mathcal{L}}{\partial A_1} &= j_1 \\
\frac{\partial \mathcal{L}}{\partial \frac{\partial A_1}{\partial t} }
&= \frac{\epsilon_0}{2} \cdot 2 \left(-\frac{\partial \phi}{\partial x_1} - \frac{\partial A_1}{\partial t}\right) (-1) \\
\frac{\partial}{\partial t} \left(\frac{\partial \mathcal{L}}{\partial \frac{\partial A_1}{\partial t} } \right)
&= \epsilon_0 \frac{\partial}{\partial t} \left(\frac{\partial \phi}{\partial x_1} + \frac{\partial A_1}{\partial t}\right)
\end{align}

\begin{align}
\frac{\partial \mathcal{L}}{\partial \frac{\partial A_1}{\partial x} }
&= - \frac{\epsilon_0 c^2}{2} \frac{\partial}{\partial \frac{\partial A_1}{\partial x}}
\left[
\left(\frac{\partial A_3}{\partial x_2} - \frac{\partial A_2}{\partial x_3} \right)^2
+ \left(\frac{\partial A_1}{\partial x_3} - \frac{\partial A_3}{\partial x_1} \right)^2
+ \left(\frac{\partial A_2}{\partial x_1} - \frac{\partial A_1}{\partial x_2} \right)^2
\right] \\
&=
- \frac{\epsilon_0 c^2}{2} 
\left(
\begin{array}{c} 
0
\\
2 \left(\frac{\partial A_2}{\partial x_1} - \frac{\partial A_1}{\partial x_2} \right) (-1)
\\
2 \left(\frac{\partial A_1}{\partial x_3} - \frac{\partial A_3}{\partial x_1} \right)
\end{array}
\right)
= 
- \epsilon_0 c^2
\left(
\begin{array}{c} 
0
\\
\left(\frac{\partial A_1}{\partial x_2} - \frac{\partial A_2}{\partial x_1} \right)
\\
\left(\frac{\partial A_1}{\partial x_3} - \frac{\partial A_3}{\partial x_1} \right)
\end{array}
\right) \\ \nonumber \\
\frac{\partial}{\partial x} \cdot \frac{\partial \mathcal{L}}{\partial \frac{\partial A_1}{\partial x} }
&= - \epsilon_0 c^2
\left( \frac{\partial^2 A_1}{{\partial x_2}^2} - \frac{\partial^2 A_2}{\partial x_2 \partial x_1}
+ \frac{\partial^2 A_1}{{\partial x_3}^2} - \frac{\partial^2 A_3}{\partial x_3 \partial x_1} \right).
\end{align}
Substituting these results into equation \eqref{ELA} results in
\begin{equation}
0= j_1 
- \epsilon_0 \frac{\partial}{\partial t} \left(\frac{\partial \phi}{\partial x_1} + \frac{\partial A_1}{\partial t} \right)
+ \epsilon_0 c^2 
\left( \frac{\partial^2 A_1}{{\partial x_2}^2} - \frac{\partial^2 A_2}{\partial x_2 \partial x_1}
+ \frac{\partial^2 A_1}{{\partial x_3}^2} - \frac{\partial^2 A_3}{\partial x_3 \partial x_1} \right).
\end{equation}
Using $\mu_0 = \frac{1}{\epsilon_0 c^2}$, this can be represented as
\begin{equation}
 - \frac{\partial^2 A_1}{{\partial x_2}^2} + \frac{\partial^2 A_2}{\partial x_2 \partial x_1}
- \frac{\partial^2 A_1}{{\partial x_3}^2} + \frac{\partial^2 A_3}{\partial x_3 \partial x_1}
= \mu_0 j_1 
+ \frac{1}{c^2} \frac{\partial}{\partial t} \left(-\frac{\partial \phi}{\partial x_1} - \frac{\partial A_1}{\partial t} \right).
\end{equation}
The right hand side of this equation is equal to the first component of the right hand side of equation \eqref{MaxwellAPhi1}.
Now we need to prove that the first component of $\nabla \times \left( \nabla \times A\right)$ equals the left hand side of this equation.
To do so we use the formula
$\nabla \times \left( \nabla \times A\right) = \nabla\left(\nabla \cdot A\right) - \Delta A$:
\begin{align}
[\nabla(\nabla \cdot A) - \Delta A]_1 &=
\frac{\partial}{\partial x_1} \left( \frac{\partial A_1}{\partial x_1} + \frac{\partial A_2}{\partial x_2} + \frac{\partial A_3}{\partial x_3} \right)
- \left( \frac{\partial^2 A_1}{{\partial x_1}^2} + \frac{\partial^2 A_1}{{\partial x_2}^2} + \frac{\partial^2 A_1}{{\partial x_3}^2} \right) \nonumber \\
&=
\frac{\partial^2 A_2}{\partial x_1 \partial x_2} + \frac{\partial^2 A_3}{\partial x_1 \partial x_3}
- \frac{\partial^2 A_1}{{\partial x_2}^2} - \frac{\partial^2 A_1}{{\partial x_3}^2}. 
\end{align}
With this, we found that the Euler-Lagrange equation for $A_1$ is equivalent to the first component of equation \eqref{MaxwellAPhi1}. To finish the proof, the last calculation must only to be repeated for the remaining Euler-Lagrange equations and components of equation \eqref{MaxwellAPhi1}.

\section{Conclusion}
The Euler-Lagrange formalism for field theories was presented as a purely mathematical framework that provides us with field equations which are invariant under any coordinate and field transformation as long as the associated Lagrangian $\mathcal{L}$ has the very simple and well defined transformation property given by equation \eqref{LagrTransform}.
Based on this mathematical result, we are well motivated to reformulate physical field equations in such a way that they become Euler-Lagrange equations.
The critical part of this reformulation is finding the Lagrangian for the field theory that we want to reformulate.

\appendix

\section{Discussion of the coordinate and field transformation defined in the beginning of section \ref{sectionInvariance}} \label{appendixDiscussionTransform}
The transformations
\begin{align}
  &x=f(\bar{x}) \;\; \text{for the spacial coordinates}\\
  &\psi=F(\bar{\psi}) \;\; \text{for the fields}
\end{align}
can be given precise meaning if we consider a particle from within two coordinate systems $\bar{T}$ and $T$.
If in $\bar{T}$ the particle's coordinates are given by $\bar{x}$, then in $T$ they are given by $x = f(\bar{x})$.

Next, we assume that in $\bar{T}$ there is a field $\bar{\psi}$ which at the particle's position $\bar{x}$ has the value $\bar{\psi}(\bar{x})$.\footnote{Of course the field may have multiple components as it would in the case of an electric field.}
We now want to know the field's value at the particle's position $x$ in $T$, which we denote by $\psi(x)$, which is where the transformation $F$ of the field comes into play.
$F$ is meant to be defined in such a way that $\psi(x)$ is given by
\footnote{A notable special case is when $F$ is the identity function.
\eqref{defFByArg} then reads $$\psi(x) = \bar{\psi}(\bar{x}) \iff \psi(f(\bar{x})) = \bar{\psi}(\bar{x}).$$
Fields that transform this way are called scalar fields.
The Higgs field is a famous example of a scalar field.}
\begin{equation} \label{defFByArg}
  \psi(x) = F(\bar{\psi}(\bar{x})) \iff \psi(f(\bar{x})) = F(\bar{\psi}(\bar{x}))
\end{equation}
The second equation allows to write the functional identity:
\begin{equation} \label{defFByFunctional}
  \psi \circ f = F \circ \bar{\psi} \;\; \text{defined on the coordinates $\bar{x}$ of $\bar{T}$}
\end{equation}
These results allow us to state two parts of section \ref{sectionInvariance} more precisely.
In fact, equations \eqref{defFByArg} and \eqref{defFByFunctional} were used in these parts:
\begin{itemize}
  \item
  In equations \eqref{lastEqToProveInvariance} and \eqref{LagrTransform} we use the derivative $\frac{\partial F(\bar{\psi})}{\partial f}$.
  For this, according to equations \eqref{defFByFunctional} and \eqref{defFByArg}, the following equation holds:
  \begin{equation}
    \frac{\partial F(\bar{\psi})}{\partial f} = \frac{\partial \psi}{\partial f} = \frac{\partial \psi}{\partial x},
  \end{equation}
  which is used to follow the equivalence of equations \eqref{lastEqToProveInvariance} and \eqref{ELGUntransformed}.
  \item
  With equation \eqref{defFByArg} it is clear that $F(x)$, which is used in equation \eqref{defineFOfx}, is well defined.
\end{itemize}

\section{Alternative to the transformation formula of multidimensional integrals} \label{TranformationFormula}

The proof of the transformation formula of multidimensional integrals is not trivial.
A heuristic explanation is given by Sterman \cite{Sterman}.
There,
\begin{equation}
\left|\det \frac{\partial f}{\partial \bar{x}} \right| =  \left|\det \frac{\partial x}{\partial \bar{x}} \right|
\end{equation}
is written as $\dd x^3 / \dd \bar{x}^3$ and is explained to be the ratio of the differentials in the transformed and untransformed coordinates.


\bibliographystyle{unsrt}
\bibliography{bibliography}

%
%
%
%
%
%
%

\end{document}